\documentclass[conference, a4paper]{IEEEtran}
\IEEEoverridecommandlockouts
\usepackage{cite}
\usepackage{amsmath,amssymb,amsfonts}
\usepackage{algorithmic}
\usepackage{graphicx}
\usepackage{textcomp}
\usepackage{xcolor}
\def\BibTeX{{\rm B\kern-.05em{\sc i\kern-.025em b}\kern-.08em
    T\kern-.1667em\lower.7ex\hbox{E}\kern-.125emX}}
    
\usepackage[normalem]{ulem}
\useunder{\uline}{\ul}{}
\usepackage{booktabs}
\usepackage{numprint}
\usepackage{acronym}

\newacro{cots}[COTS]{Commercial Off The Shelf}   
\newacro{it}[IT]{Information Technology}  
\newacro{hmi}[HMI]{Human Machine Interface}
\newacro{plc}[PLC]{Programmable Logic Controller}  
\newacro{cnc}[C\&C]{Command \& Control}  
\newacro{ics}[ICS]{Industrial Control System} 
\newacro{scada}[SCADA]{Supervisory Control And Data Acquisition}  
\newacro{rnn}[RNN]{Recurrent Neural Network}  
\newacro{lstm}[\textit{LSTM}]{\textit{Long Short-Term Memory}}

\begin{document}

\title{Devil in the Detail:\\Attack Scenarios in Industrial Applications
\thanks{This is a preprint of a publication accepted at the 2019 IEEE Workshop on the Internet of SafeThings.
Please cite as: S. D. Duque Anton, A. Hafner, H. D. Schotten,
``Devil in the Detail: Attack Scenarios in Industrial Applications,'' in \textit{2019 IEEE Security and Privacy Workshops (SPW)}. IEEE, 2019}
}

\author{\IEEEauthorblockN{1\textsuperscript{st} Simon D. Duque Anton}
\IEEEauthorblockA{\textit{Intelligent Networks Research Group} \\
\textit{German Research Center for AI}\\
67663 Kaiserslautern, Germany\\
Simon.Duque\_Anton@dfki.de}
\and
\IEEEauthorblockN{2\textsuperscript{nd} Alexander Hafner}
\IEEEauthorblockA{\textit{Intelligent Networks Research Group} \\
\textit{German Research Center for AI}\\
67663 Kaiserslautern, Germany\\
Alexander.Hafner@dfki.de}
\and
\IEEEauthorblockN{3\textsuperscript{rd} Hans Dieter Schotten}
\IEEEauthorblockA{\textit{Intelligent Networks Research Group} \\
\textit{German Research Center for AI}\\
67663 Kaiserslautern, Germany\\
Hans\_Dieter.Schotten@dfki.de}

}

\maketitle

\begin{abstract}
In the past years,
industrial networks have become increasingly interconnected and opened to private or public networks.
This leads to an increase in efficiency and manageability,
but also increases the attack surface.
Industrial networks often consist of legacy systems that have not been designed with security in mind.
In the last decade,
an increase in attacks on cyber-physical systems was observed,
with drastic consequences on the physical work.
In this work,
attack vectors on industrial networks are categorised.
A real-world process is simulated,
attacks are then introduced.
Finally,
two machine learning-based methods for time series anomaly detection are employed to detect the attacks.
\textit{Matrix Profiles} are employed more successfully than a predictor \textit{Long Short-Term Memory} network,
a class of neural networks.
\end{abstract}

\begin{IEEEkeywords}
Cyber Security, Time Series, Machine Learning, Neural Networks, Industrial Control Systems
\end{IEEEkeywords}

\section{Introduction}
For decades,
the industrial domain has been deemed secure due to two reasons:
First,
the physical separation of networks.
Second,
each network was created in an application specific fashion,
rendering it extremely difficult for an attacker to exploit it~\cite{Igure.2006}.
However,
the fourth industrial revolution introduced novel use cases that build on interconnectivity and embedded intelligence~\cite{3gpp2017,  Plaga.2019}.
While increasing productivity and flexibility and decreasing operational cost and effort,
new attack vectors are introduced to industrial systems as well.
An increase in attacks on industrial environments can be detected~\cite{Duque_Anton.2017a}.
While industrial networks have been unique in their applications specific nature,
the establishment of \ac{cots} hard- and software introduces standardised modules.
This makes set up and maintenance much easier,
but also drastically increases the effect of vulnerabilities in one of the modules.
In order to tackle these problems,
cyber security measures have been adapted to industrial scenarios,
such as firewalls,
anti virus software and intrusion detection tools.
However,
the characteristics of industrial networks differ from those of home and office networks,
motivating the need for adaption of those tools.
A deep understanding of these characteristics is required in order to effectively protect industrial networks.
In this work,
an overview of possible attacks for industrial networks is provided.
Attack vectors are analysed and categorised,
with an emphasis on industrial network protocols.
Furthermore,
the simulation of a real-world scenario is presented,
as well as attacks on this scenario.
The remainder of this work is structured as follows:
In Section~\ref{sec:sota},
related work is presented.
A systematic categorisation of attack scenarios is provided in Section~\ref{sec:attack_scenarios}.
The simulated process and the implementation of attacks is described in Section~\ref{sec:implementation} and evaluated in Section~\ref{sec:evaluation}.
Finally,
the findings are discussed in Section~\ref{sec:discussion}.

\section{Related Work}
\label{sec:sota}
In this section,
related work on classification of industrial cyber attacks is presented.
Furthermore,
it is grouped with respect to the scope that is addressed by the work in Table~\ref{tab:sota}.
\textit{Cherdantseva et al.} survey existing risk assessment methods and evaluate their usefulnes with respect to \ac{scada} scenarios~\cite{Cherdantseva.2016}.
\textit{Gao and Morris} discuss the detection of cyber attacks~\cite{Gao.2014}.
They focus on signature-based detection for \textit{Modbus}-based communication.
In order to evaluate the intrusion detection and to classify it,
possible attacks are grouped.
A more thorough analysis of attacks on \acp{ics} is performed by \textit{Morris and Gao} as well~\cite{Morris.2013}.
\textit{Zhu et al.} provide an overview of cyber attacks while considering many dimensions~\cite{Zhu.2011}.
They compare industrial cyber security to classic IT security.
Furthermore,
they consider the security objectives of industrial applications and ways they can be attacked.
Finally,
they present specific attacks on different attack surfaces of an industrial environment.
In another work, 
\textit{Zhu and Sastry} create a taxonomy for \ac{scada}-specific attacks\cite{Zhu.2010}.
They present types of attacks and discuss countermeasures.
\textit{Fernandez et al.}  discuss the development of secure \ac{scada} systems~\cite{Fernandez.2010}.
In doing so,
attacks on industrial systems are evaluated with respect to their attack vector.
\textit{Fovino et al.} discuss the effects of \ac{scada} attacks on infrastructure~\cite{Fovino.2009}.
They first assess the potential damages to eventually discuss potential attack types.
\textit{Ten et al.} present a vulnerability assessment of \ac{scada} systems~\cite{Ten.2008}. 
They consider the increasing dependency of industrial and office \ac{it}. 
Furthermore,
they classify attacks according to their type to model and evaluate attack scenarios in using attack trees in an earlier work~\cite{Ten.2007}.
\textit{Cai et al.} analyse the development of \ac{scada} systems,
their applications and threats~\cite{Cai.2008}.
Furthermore, 
they discuss standards and guidelines for protecting such systems.
\textit{Igure et al.} discuss \ac{scada} security~\cite{Igure.2006}.
They analyse attacks, 
categorise them and extract research challenges.
Furthermore,
standardisation efforts are addressed.
The summary of addressed topics is shown in Table~\ref{tab:sota}.

\begin{table}[h!]
\renewcommand{\arraystretch}{1.3}
\caption{Research Topics Covered by the Individual Works}
\label{tab:sota}
\centering
\scriptsize
\begin{tabular}{l  l}
\toprule
\textbf{Subject Covered} & \textbf{Research Work} \\
Risk Assessment & \cite{Cherdantseva.2016, Fovino.2009, Ten.2008, Ten.2007} \\
Industrial vs Home- and Office  \ac{it} & \cite{Zhu.2011} \\
Attack Vectors & \cite{Fernandez.2010} \\
Security Objectives & \cite{Zhu.2011, Igure.2006} \\
Types of Attacks & \cite{Gao.2014, Morris.2013, Zhu.2011, Zhu.2010, Fovino.2009, Ten.2008, Ten.2007, Igure.2006} \\
Standards and Guidelines & \cite{Cai.2008} \\
Applications in \ac{scada} & \cite{Cai.2008} \\
Taxonomy & \cite{Zhu.2010, Igure.2006} \\
Intrusion Detection & \cite{Gao.2014, Zhu.2010} \\
\bottomrule
\end{tabular}
\end{table}

Most research is done regarding the types of attacks,
i.e. the way an attacker will influence the systems or networks.
Risk assessment is a widely regarded topic as well.
In risk assessment,
the effects of an attack are discussed in a formal manner.
The remaining topics are more specific and only addressed by one or two singular works.

\section{Industrial Attacks}
\label{sec:attack_scenarios}
In this section,
possible ways for an attacker to  break into industrial applications are discussed.
This is done by looking at past attacks on industrial networks that have extensively been discussed.
\textit{Stuxnet}, 
the attack that came to attention first,
has been widely discussed~\cite{Dragos.2016, Langner.2013, Virvilis.2013, Lindsay.2013},
but also the lesser known successors,
such as \textit{Duqu}~\cite{Virvilis.2013},
\textit{Industroyer/Crashoverride}~\cite{Cherepanov.2017, Dragos.2016},
\textit{Flame}~\cite{Virvilis.2013},
\textit{BlackEnergy}~\cite{Cherepanov.2017, Dragos.2016},
\textit{Havex}~\cite{Dragos.2016} and \textit{Red October}~\cite{Virvilis.2013} have received attention.
An assessment of attack vectors for industrial companies is done by \textit{Positive Technologies}~\cite{Positive_Technologies.2018}.
They evaluate points of entry and propagation methods in a general fashion which,
however,
is in accordance to the above-mentioned malware-specific analyses. 
The first step in attacking industrial environments is commonly the breach of the perimeter.
Even though there are occasions where industrial networks are directly connected to the Internet~\cite{Duque_Anton.2017a},
they are commonly separated from public networks.
This is an important recommendation in securing industrial networks~\cite{Positive_Technologies.2018},
especially since many industrial network protocols do not contain means for authentication or encryption.
This allows easy propagation and participation in communication for an attacker once the network is accessible.
If the production network is not reachable from the outside,
the corporate network needs to be breached first.
According to \textit{Positive Technologies},
73\% of the corporate systems they tested had insufficient protection of their perimeter~\cite{Positive_Technologies.2018}.
Another common attack vector is the human user.
Allegedly the \textit{Stuxnet} attack has breached the perimeter by means of a thumb drive that was carelessly used~\cite{Lindsay.2013}.
After breaking the perimeter,
the \ac{ics} or the field devices respectively have to be taken over.
The analysed malware that was tailor-made for industrial targets used properties or vulnerabilities characteristic to the industrial environment they were designed for.
Some malware could stay undetected for long periods of time,
at least partly due to missing or insufficient security procedures for industrial networks.
Implementing robust security for critical parts of production networks is one of the major take aways.
Most industrial malware consists of several modules:
\begin{itemize}
\item Backdoor,
\item loader module, and
\item wiper module
\end{itemize}
The backdoor allows for communication with \ac{cnc} servers.
Coupled to the backdoor is the loader module that is tasked with uploading the malware modules to perform certain attacks.
And lastly,
most industrial malware contains a module for wiping the traces of its existence from the infected system.
Breaking the perimeter has proven to be possible most of the time.
The difficulty of industrial malware lies in the profound knowledge the malware authors needs to have about the targeted systems.
This goes for the architecture of the infrastructure as well as for the protocols and devices used.
Most devices used in industrial applications are \ac{cots} products and can thus be obtained for vulnerability analysis,
so that exploits can be written and re-used.
However,
to successfully break a process by abusing system parameters,
the intent of devices as well as the structure of the process needs to be known.
These attacks are hardest to detect,
as the attacker can conceal them as irregularities or normal behaviour.
The attacks that are implemented and evaluated in this work are such attacks.
They are based on the assumption of a successful breach of perimeter and take over of a \ac{plc}  which subsequently shows malicious behaviour.\\
In summary,
any attack of an industrial application first needs to break the perimeter.
Then it needs to move laterally towards the control system or target device.
Finally,
the malicious intent has to be carried out.
During each of these steps,
the attack can be discovered by different means.
Breaking the perimeter should be observed by \ac{it}-based security means.
Lateral movement is in the domain of \ac{siem}-systems.
Detecting attacks in the context of an industrial process is the final method to discover misbehaviour.

\section{Process Environment and Attack Scenarios}
\label{sec:implementation}
In this section,
the process under investigation as well as the implemented attacks are discussed.
First,
the real-world process is described and transferred to a simulation.
After that, 
the attack scenarios and their implementations are presented.

\subsection{Process Environment}
\begin{figure}
  \includegraphics[width=\linewidth]{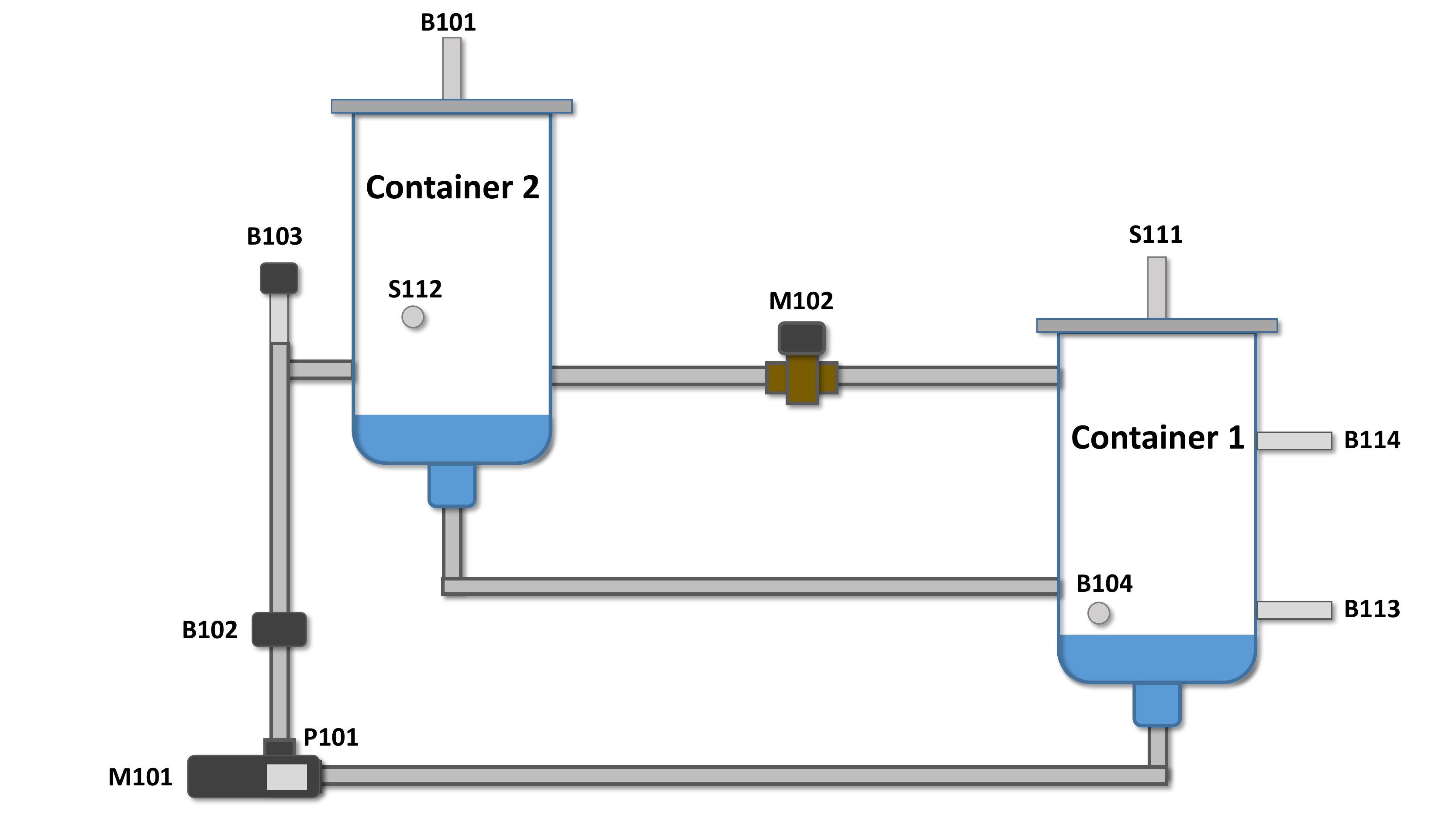}
  \caption{Schematic Overview of the Process Environment}
  \label{fig:process}
\end{figure}

\begin{figure*}
  \includegraphics[width=0.97\linewidth]{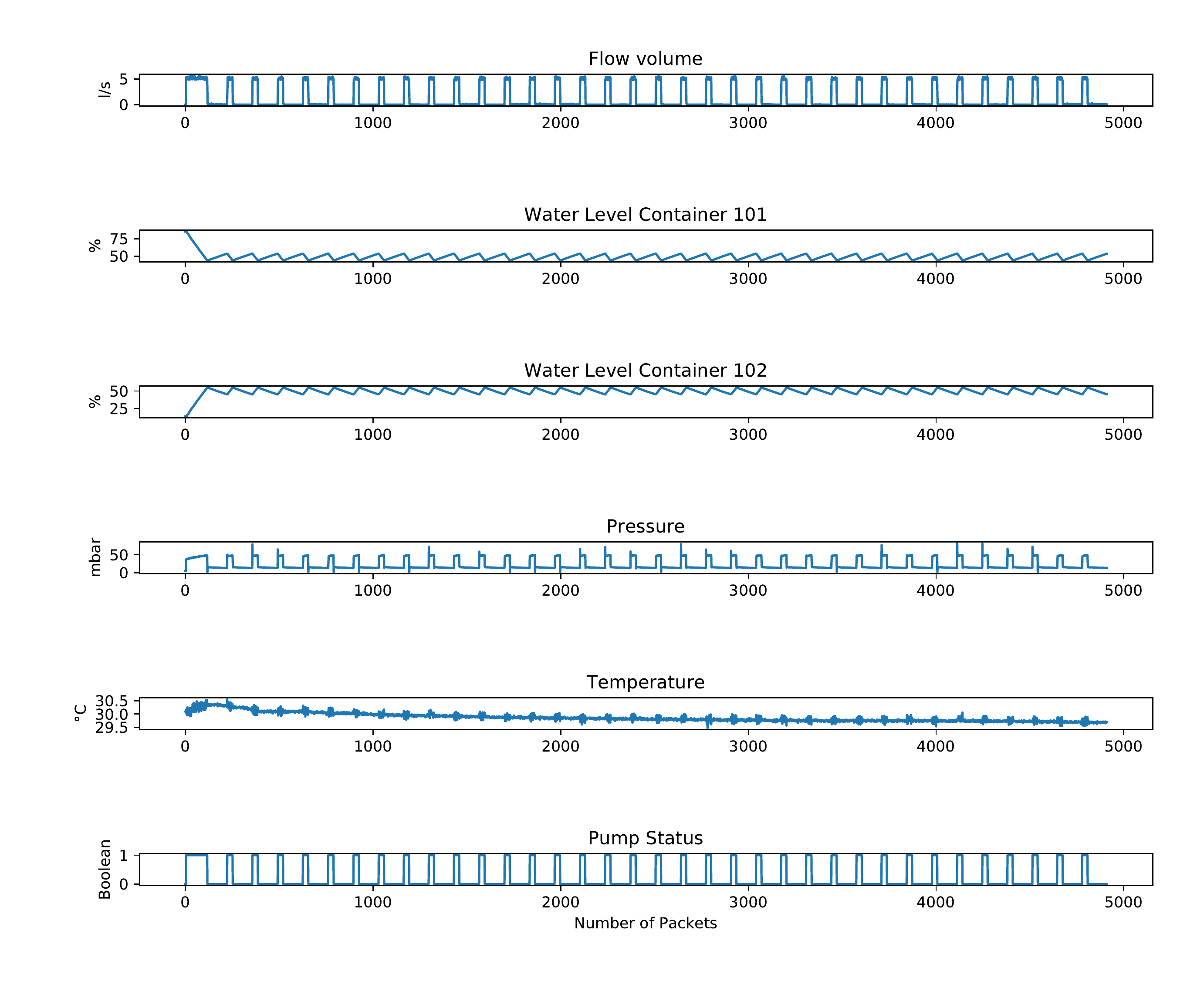}
  \caption{Normal Process Behaviour}
  \label{fig:normal_traffic}
\end{figure*}

The process this work is based upon has been used to generate data for industrial intrusion detection already~\cite{Duque_Anton.2019a}.
It is shown in Figure~\ref{fig:process}.
The process environment consists of two water containers,
\textit{Container 1} and \textit{Container 2}.
Water is pumped with pump \textit{P101},
driven by a DC motor \textit{M101} from \textit{Container 1} to \textit{Container 2} until a threshold is reached.
The water level is measured with different sensors,
 \textit{S111} and \textit{S112},
 as well as capacitive sensors \textit{B113} and \textit{B114}.
 Additionally,
 A vane sensor measuring the flow of liquid between \textit{Container 1} and \textit{Container 2},
 \textit{B102},
 and a \textit{PT100} temperature sensor,
 \textit{B104},
 are used.
 To release water from \textit{Container 2},
 a solenoid valve,
 \textit{M102},
 is employed.
 An exemplary behaviour of this process is shown as a time-series in Figure~\ref{fig:normal_traffic}.
A selection of process parameters,
all of them sensor outputs,
during normal operation is shown. 
This operation has been performed on real implementation of the scenario that was used to create the simulation analysed in this work.
For this work,
the environment described above has been extended to five instances.
They are simulated with real-world hardware,
i.e. \textit{Siemens S7-1500} \acp{plc} and \textit{PiXtend} extension boards for \textit{Raspberry Pis}.
Five \textit{Raspberry Pis} with a \textit{PiXtend}-board each are used to simulate the process,
controlled by a \ac{plc} each.
The process information is collected on a central \ac{hmi}.
In order to obtain realistic data,
the simulation has been developed to mimic the real scenario as good as possible.
For communication,
\textit{OPC UA}~\cite{IEC.2016} is used.
It provides encrypted,
authenticated,
easy and platform-independent communication and consists of an information model including communication capabilities.
In this scenario,
a master-slave concept is followed with regular polling of the devices by the \ac{hmi}.

\subsection{Attack Scenarios}
\label{ssec:attack_scen}
\begin{figure}
  \includegraphics[width=\linewidth]{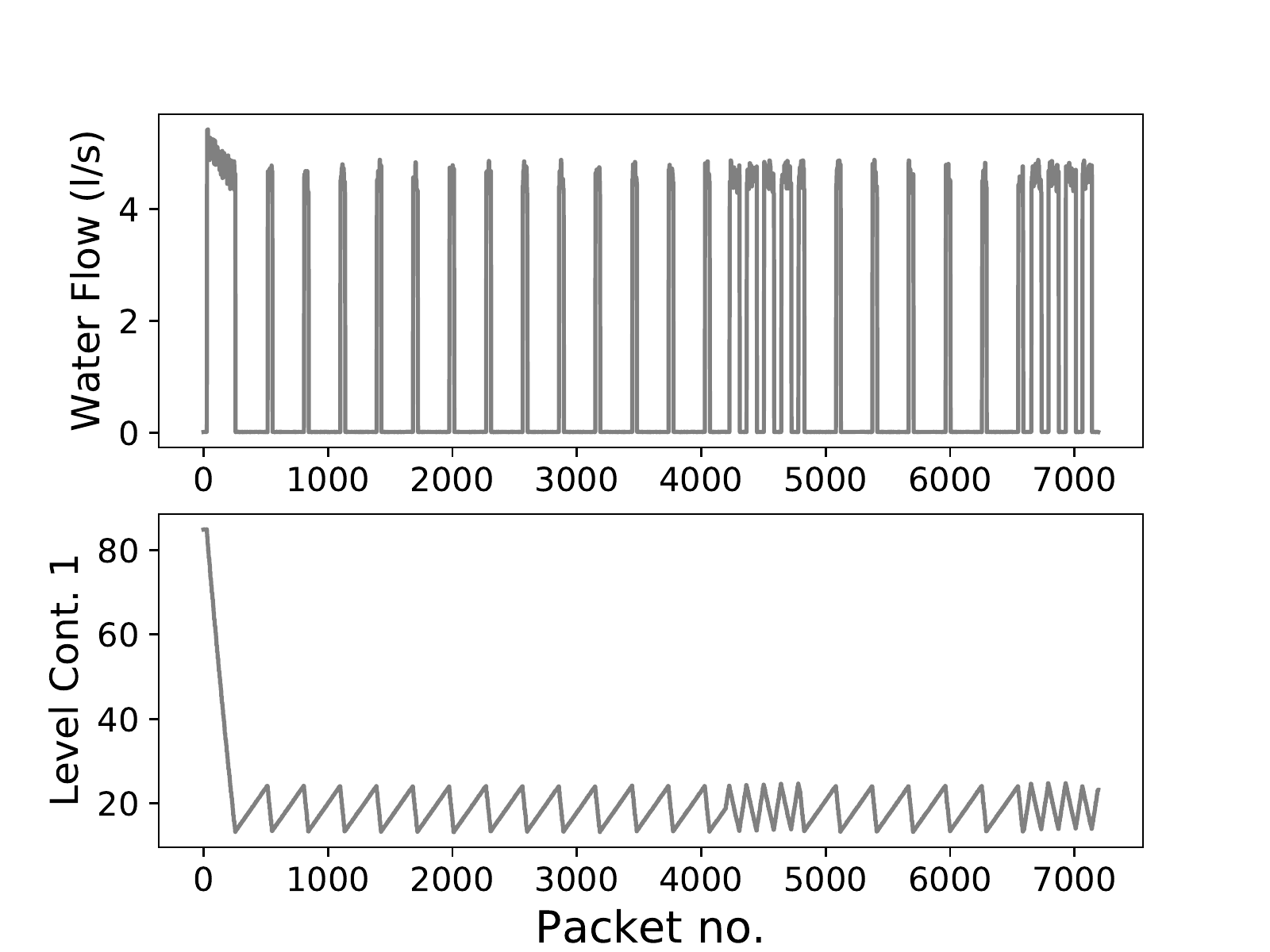}
  \caption{Malicious Process Behaviour}
  \label{fig:mal_proc}
\end{figure}

Two scenarios have been implemented and evaluated in this work.
They are loosely coupled to the categorisation of \textit{Morris and Gao}~\cite{Morris.2013}.
In this work,
all attacks are a kind of \textit{Response and Measurement Injection Attack}.
In creating the data set for evaluation,
one of the five \acp{plc} shows malicious behaviour for five minutes after 15 minutes of normal operation for each attack.
The use case is an attacker having breached the perimeter, 
bridged the air gap and used well-engineered malicious code to disrupt the process.
The aim of this scenario is to detect malicious behaviour on field level.
This area is currently not well-developed,
intrusion detection on field level is a growing field with a short history.
The behaviour regarding flow and water level of \textit{Container 1} of the malicious process is shown in Figure~\ref{fig:mal_proc}.
Since a specific application use case is discussed in this work,
the attacks are domain-specific.
However,
as discussed in Section~\ref{sec:attack_scenarios},
the general concept of this kind of attack can be generalised to most processing units.

\subsubsection{Open Valve Attack}
In the first attack scenario,
the valve \textit{M102} is opened even though water is pumped from \textit{Container 1} to \textit{Container 2}.
This leads to an increased time it takes for \textit{Container 2} to be filled up to the desired level.
The \ac{hmi} indicates the valve as open.
This attack starts at packet 4,200 and ends around packet 4,800 in Figure~\ref{fig:mal_proc}.
As in the process,
the valve is not supposed to be opened,
this is an identifier of the attack,
making it trivial to detect.
In order to better evaluate the methods presented in this work,
it is not used as an input variable.

\subsubsection{Stealth Attack}
The second attack scenario is implemented in a stealthier fashion.
As in attack scenario 1,
the valve is opened invalidly.
However,
the sensor still indicates a closed valve.
This leads to an unexpected decrease in filling speed of \textit{Container 2} and an increased emptying once the container is filled.
This attack starts at packet 6,500 and ends at the end of the trace in Figure~\ref{fig:mal_proc}.

\section{Evaluation}
\label{sec:evaluation}
In this section,
the methods to detect the discussed attacks are presented and evaluated on the data set.
As input values for the anomaly detection,
the water flow as well as the water level of \textit{Container 1} are used.
They could easily be extended,
but proved to be the most expressive variables.

\subsection{Matrix Profiles}
\begin{figure}
  \includegraphics[width=\linewidth]{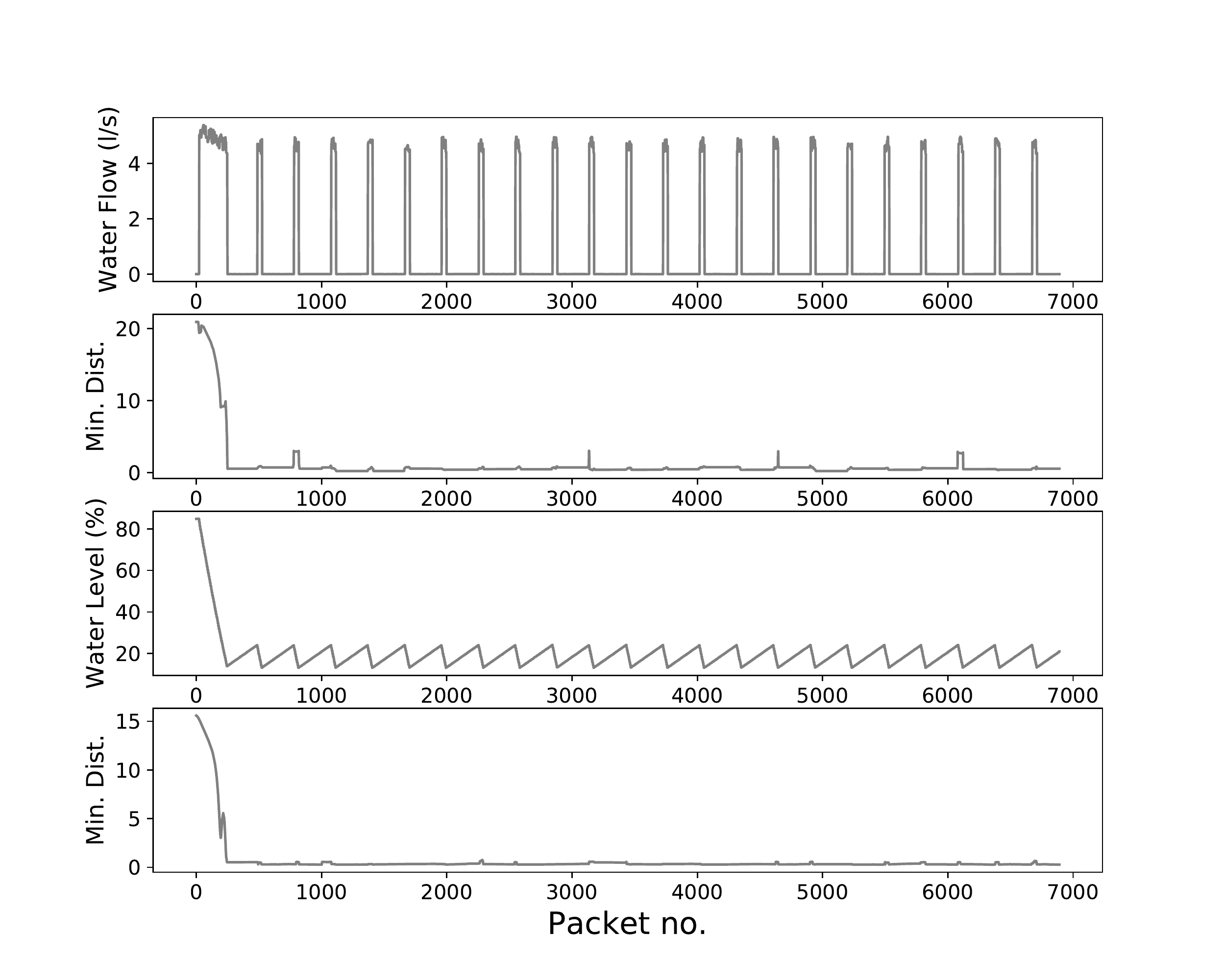}
  \caption{Time Series and \textit{Matrix Profiles} of Normal Behaviour}
  \label{fig:ts_normal}
\end{figure}

\begin{figure}
  \includegraphics[width=\linewidth]{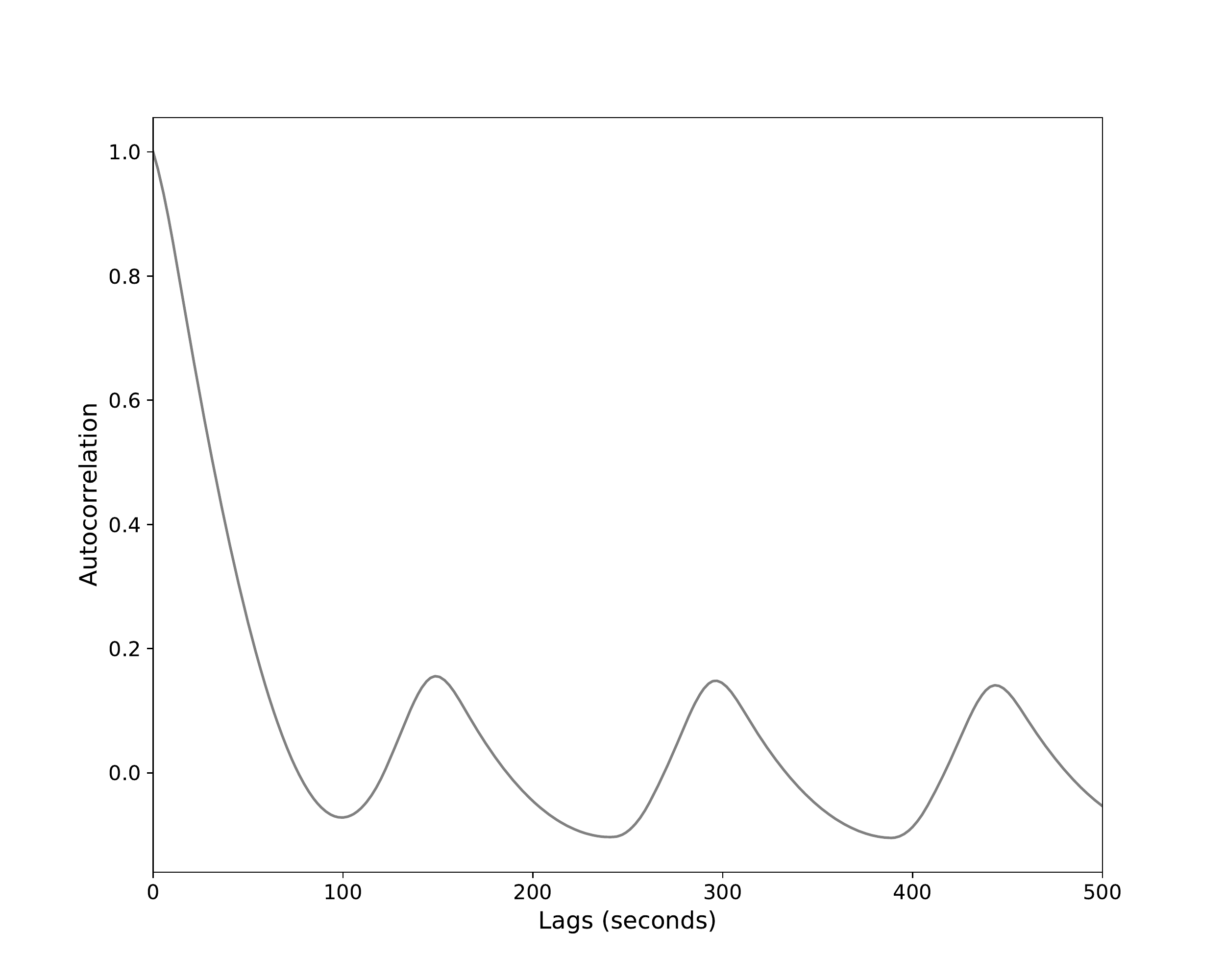}
  \caption{Auto-correlation of Normal Time Series}
  \label{fig:acf}
\end{figure}

\begin{figure}
  \includegraphics[width=\linewidth]{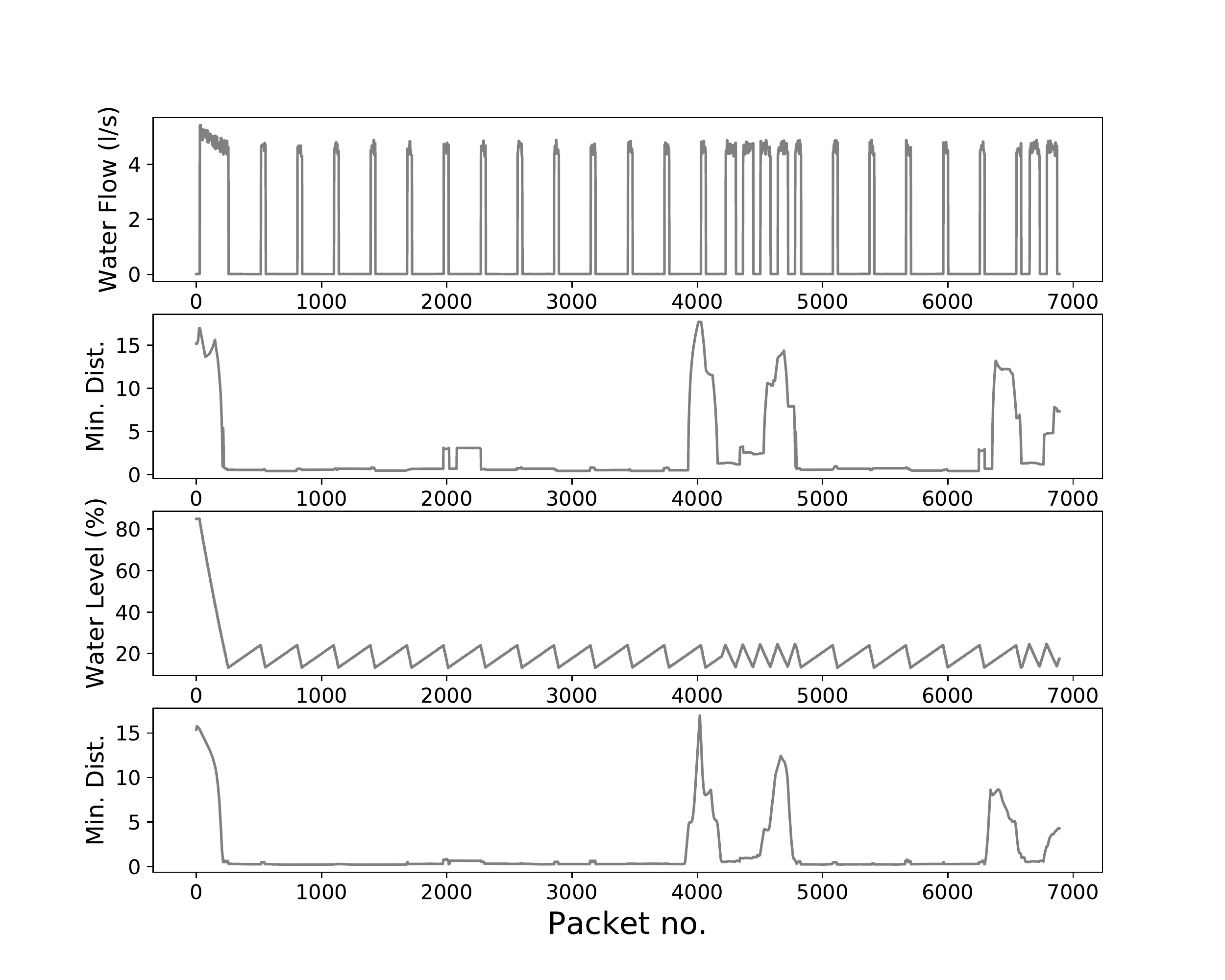}
  \caption{Time Series and \textit{Matrix Profiles} of Malicious Behaviour}
  \label{fig:ts_mal}
\end{figure}

Time series-based anomaly detection has proved to be highly effective in industrial intrusion detection~\cite{Duque_Anton.2018c}.
As the process is expected to produce regular sensors and actuator values,
deviations of a time series representation of those values should be detectable.
In this work,
\textit{Matrix Profiles} are used to analyse the data sets.
\textit{Matrix Profiles} were introduced by \textit{Yeh et al.} in 2016 and provide a mean to determine the similarity of sequences in a time series to other sequences~\cite{Yeh.2016a}.
In order to employ \textit{Matrix Profiles},
only one hyper-parameter needs to be set,
the window size $m$.
It is robust against changes and provides sensible results for a variety of lengths.
However,
each time window needs to contain a set of values that has a standard deviation that is not zero.
Thus,
$m$ needs to be chosen in a way that no window in the water flow values only contains zero flow,
as shown in Figure~\ref{fig:ts_normal}.
In this figure,
the water flow,
as well as the water level of \textit{Container 1},
are shown in combination with their respective \textit{Matrix Profiles}.
The \textit{Matrix Profile} determines the minimal distance of any windowed sequence of length $m$ from any other sequence of length $m$.
In terms of anomaly detection,
a high minimal distance represents an outlier,
as the corresponding window does not look like any other.
In Figure~\ref{fig:ts_normal},
the normal behaviour of the process is shown,
with $m$ as 300.
Even though $m$ proved to be robust in the evaluation,
auto-correlation~\cite{Gubner.2006} was employed in order to find a sensible value.
The auto-correlation function of the normal process is shown in Figure~\ref{fig:acf}.
The peak at around 150 seconds indicated periodic behaviour.
In our experiments,
two measurements per second were performed,
thus a window size of 300 packets was chosen as $m$.
In Figure~\ref{fig:ts_normal},
the \textit{Matrix Profiles},
named \textit{Min. Dist.},
are small,
except for the beginning.
The settling of the process is a unique event,
thus the high minimal distance.
The process was monitored on one of the \acp{plc} that did not exhibit malicious behaviour.
This was introduced to another \ac{plc} and resulted in the behaviour shown in Figure~\ref{fig:ts_mal}.
Both attacks show significant peaks in the minimal distances around both attacks as described in Subsection~\ref{ssec:attack_scen}.
It is noteworthy that the transitions from normal to malicious behaviour,
and vice versa,
are the events in the time series that are unique and thus result in an increased minimal distance.
If an attack has a characteristic signature that is repeated more than once,
it is not detected as an anomaly anymore,
as another instance with the same characteristic is found.
To counter this effect,
\textit{Matrix Profiles} can be adapted in a way that they are employed continuously.
This shows promising results~\cite{Duque_Anton.2018c}.
This approach indicates any change in behaviour.
To mitigate disruptions due to alerts,
natural changes in processes can be integrated into \textit{Matrix Profiles} with an extension~\cite{Zhu.2017}.
Furthermore,
\textit{Matrix Profiles} can be used to analyse meta data,
providing good results as well~\cite{Duque_Anton.2018c}.

\subsection{Long Short-Term Memory}
\begin{figure}
  \includegraphics[width=\linewidth]{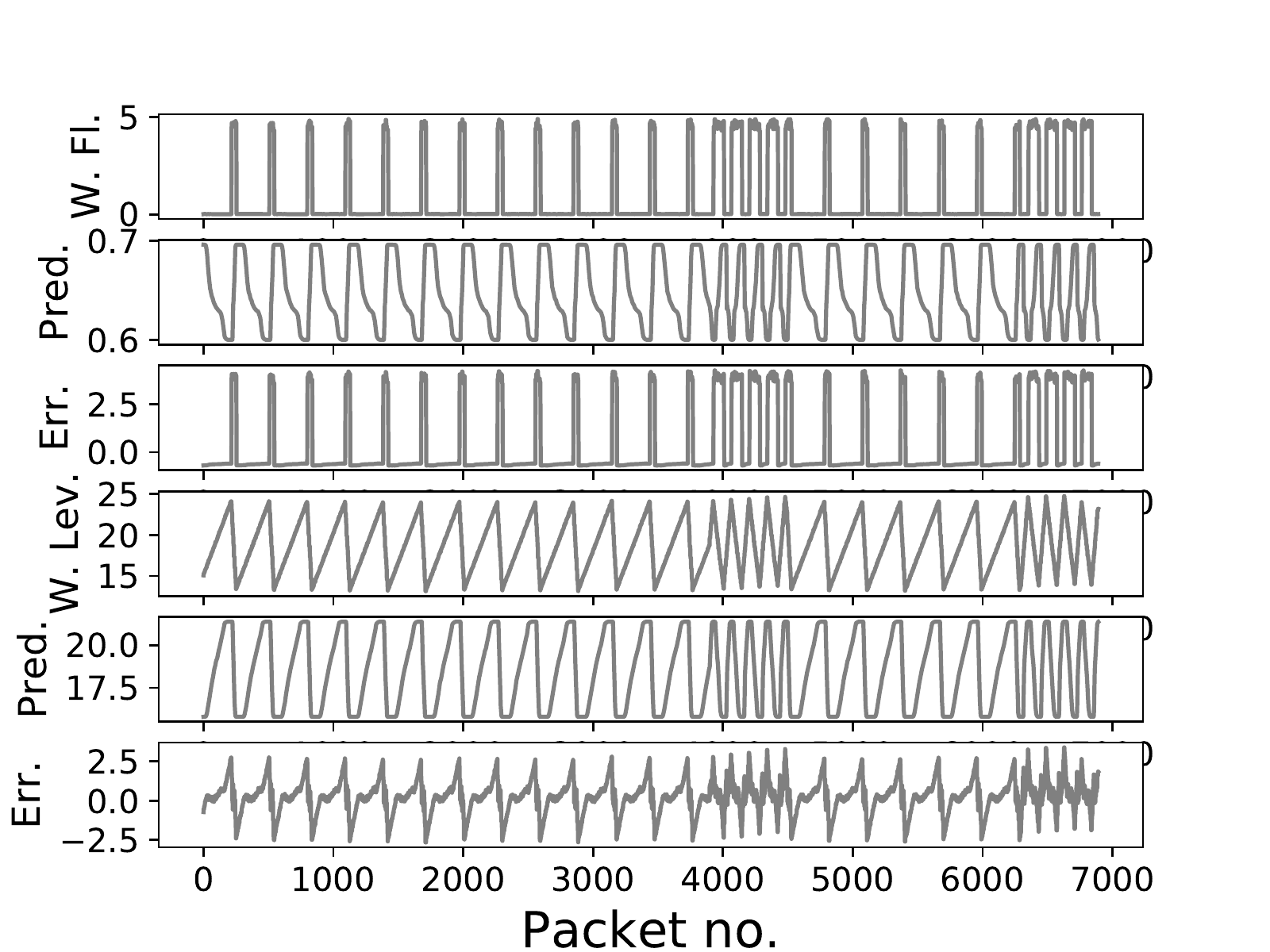}
  \caption{Time Series, Predictions and Errors Based on \acp{lstm}}
  \label{fig:lstm}
\end{figure}

Many types of \acp{rnn} tend to neglect long-term dependencies in the decision making.
In order to keep such information,
\textit{Hochreiter and Schmidhuber} proposed a novel kind of \acp{rnn} in 1997 to overcome the vanishing gradient problem~\cite{Hochreiter.1997}.
This kind of \ac{rnn} is called \ac{lstm}.
In this work,
an \ac{lstm} with an input layer consisting of 350 units, 
two hidden layers with 350 and 250 units respectively and a dense output-layer is employed.
The input length is 300 as this is the minimal periodicity of the data.
The learning rate was set to 0.001.
An hour of process activity was used to train it in 25 iterations.
No attacks were contained in the data.
After that,
the hour of process activity conducted by the infected \ac{plc} was used as the testing data set.
In order to monitor anomalousness,
a value was predicted by the neural network and compared to the real value.
The distance between those values was calculated,
a high value indicating an anomalous instance.
The result of the \ac{lstm} is shown in Figure~\ref{fig:lstm}.
The first row shows the real values of water flow,
the second row the predicted values and the third row the absolute error.
The fourth row shows the real values of the water \textit{Container 1},
the second row the predicted values and the third row the absolute error.
It can be seen that the \ac{lstm} closely follows the process behaviour.
Unfortunately,
this also includes the attacks.
They are predicted as part of the process by the \ac{lstm},
making detection of the attacks difficult. 
Only the frequency of the periodic error behaviour changes,
however,
values that clearly indicate attacks would enhance the detection probability.

\section{Discussion}
\label{sec:discussion}
In this work,
we discussed the attack scenarios in industrial environments.
From these scenarios,
a use case was derived and implemented.
After that, 
attack scenarios where introduced to the scenario.
Two time series-based anomaly detection methods were employed to detect the attacks.
\textit{Matrix Profiles} performed satisfactorily,
detecting the attacks easily.
Only one robust hyper-parameter and no supervised training make it easy to use and transfer between application domains.
The \ac{lstm} approach did not work well,
it predicted the attack behaviour as well as the normal behaviour.
This behaviour can derive from over-fitting,
as regular time series have a tendency to teach neural networks to learn certain patterns,
but not to generalise.
Context information~\cite{Duque_Anton.2017c} or methods of machine learning-based classification~\cite{Duque_Anton.2018b} might address the issue as well.
\subsection{An Epilogue on Sophisticated Industrial Attacks}
One of the major features of sophisticated industrial attacks such as \textit{Stuxnet} is the masquerading of any indicators for misbehaviour.
However,
if no trace of malicious behaviour is simulated, 
it simply cannot be detected.
For the sake of clarity in this work,
only attacks with distinctive characteristics were used,
so that detection was possible.
After attacks such as \textit{Stuxnet} propagated into the industrial domain,
side-channel detection,
e.g. acoustic signals,
would be required if standard field busses were used.
\textit{Langner} claims that any engineer with experience in the area would have told something was amiss easily by the sound of the turbines.
Unfortunately,
such side-channels are hard to simulate.
However,
there are works creating data sets of real-world applications including side-channel sensor measurements so that they can be used to detect attacks~\cite{Duque_Anton.2019a}.

\section*{Acknowledgment}
This work has been supported by the Federal Ministry of
Education and Research of the Federal Republic of Germany
(Foerderkennzeichen 16KIS0932, IUNO Insec). The authors alone
are responsible for the content of the paper.

%The preferred spelling of the word ``acknowledgment'' in America is without 
%an ``e'' after the ``g''. Avoid the stilted expression ``one of us (R. B. 
%G.) thanks $\ldots$''. Instead, try ``R. B. G. thanks$\ldots$''. Put sponsor 
%acknowledgments in the unnumbered footnote on the first page.

\bibliographystyle{IEEEtran}
\bibliography{literature}

\end{document}